\documentclass[pra,twocolumn]{revtex4}
\usepackage{amsmath}                          
\usepackage{amsfonts}                         
\usepackage{amssymb}                          
\usepackage{amscd}                            
\usepackage{amsthm}                           

\usepackage[utf8]{inputenc}

\usepackage{graphicx}                         
\usepackage{epstopdf}                         
\usepackage[colorlinks=true,
            urlcolor=blue,
            citecolor=blue,
            linkcolor=black,
            pdfstartview=FitH,
            pdfpagemode=None]{hyperref}

\usepackage{datetime}                         

\newcommand{\lsim}{\mathrel{\raise.3ex\hbox{$<$\kern-.75em\lower1ex\hbox{$\sim$}}}}
\newcommand{\gsim}{\mathrel{\raise.3ex\hbox{$>$\kern-.75em\lower1ex\hbox{$\sim$}}}}

\def\QECCnk[[#1,#2]]{[\![#1, #2]\!]}

\def\QECCnkq[[#1,#2,#3]]{[\![#1, #2]\!]_{#3}^{\vphantom{T}}}

\def\QECCnkd[[#1,#2,#3]]{[\![#1, #2, #3]\!]}

\def\QECCnkdq[[#1,#2,#3,#4]]{[\![#1, #2, #3]\!]_{#4}^{\vphantom{T}}}

\def\QECCnkgd[[#1,#2,#3,#4]]{[\![#1, #2, #3, #4]\!]}

\def\QECCnkgdq[[#1,#2,#3,#4,#5]]{[\![#1, #2, #3, #4]\!]_{#5}^{\vphantom{T}}}

\def\QECCnkdc[[#1,#2,#3,#4]]{[\![#1, #2, #3; #4]\!]}

\def\QECCnkdcq[[#1,#2,#3,#4,#5]]{[\![#1, #2, #3; #4]\!]_{#5}^{\vphantom{T}}}

\def\QECCnkgdcq[[#1,#2,#3,#4,#5,#6]]{%
  [\![#1, #2, #3, #4; #5]\!]_{#6}^{\vphantom{T}}}

\newcommand{\bigO}{{\cal O}}

\def\openone{\leavevmode\hbox{\small1\normalsize\kern-.33em1}}

\long\def\symbolfootnote[#1]#2{\begingroup%
\def\thefootnote{\fnsymbol{footnote}}\footnote[#1]{#2}\endgroup}

\input{Qcircuit.tex}

\theoremstyle{definition}  

\begin{document}


%
\title{On the Power of Reusable Magic States}

\author{Jonas T. \surname{Anderson}}
\email[]{jander10@unm.edu}
\affiliation{Center for Quantum Information and Control,
             University of New Mexico,
             Albuquerque, NM, 87131, USA}

\date{\today, \currenttime}


\begin{abstract}
In this paper we study reusable magic states. These states are a special subset of the standard magic states. Once distilled, reusable magic states can be used repeatedly to apply some unitary U. Given this property, reusable magic states have the potential to greatly lower qubit and gate overhead in fault-tolerant quantum computation. While these states are promising, we provide a strong argument for their limited computational power. Specifically, we show that if reusable magic states can be used to apply non-Clifford unitaries, then we can exploit them to efficiently simulate poly-sized quantum circuits on a classical computer.
\end{abstract}

%
\maketitle


%
\section{Introduction}

Magic states were introduced by Bravyi and Kitaev \cite{Bravyi:2005a} as a way to implement logical gates that were not available as transversal gates in an error-correcting code. Their idea was as follows: first prepare many initial magic states, then use these to create an encoded magic state. This encoding procedure will introduce noise, and the encoded magic state will only be close to the desired state. We repeat this process to obtain many noisy encoded magic states. We then put these through many rounds of a distillation protocol. This eventually produces an encoded magic state of the desired fidelity. Finally, we use gate teleportation to apply the gate corresponding to the magic state to our encoded state. 

The procedure described above is the canonical way of completing a universal gate set. In fact, Eastin and Knill \cite{Eastin:2008a} proved that universal and transversal gate sets do not exist for any quantum code, thereby making magic state distillation not only a convenience, but a necessity. Note that there exist other possible ways of completing a universal gate set such as braiding of anyons \cite{Kitaev:1997a} or Dehn twists \cite{Koenig:2010a}, but these will not be discussed here.   

Once we have decided to use magic states, it is important to focus on lowering the immense overhead associated with them. This can be done in a variety of ways which will be discussed in turn below. 
\begin{figure}[!htb]
\center{
\includegraphics[width=1.0\columnwidth]{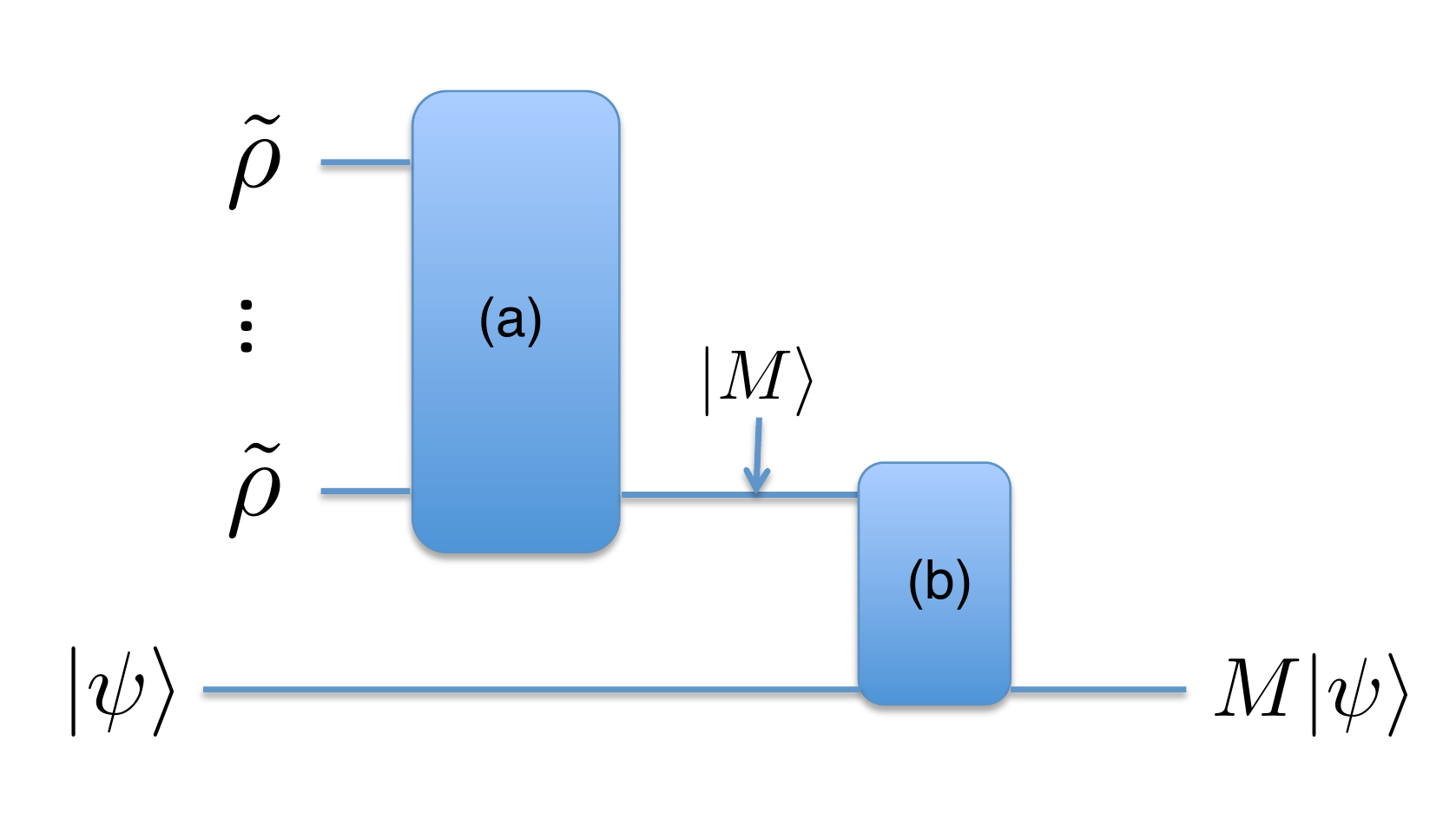}
} 
\caption{Unencoded magic protocol. (a) Distillation. (b) Gate Teleportation.}
\label{fig:protocolunencoded}
\end{figure}
%

%
%
%
%
%
%
%

\begin{description}
\item[(1)] We can choose a code with many transversal gates available. As mentioned above, it is not possible to have a universal and transversal gate set, but we can come close. For example, the 15-qubit Reed-Muller code \cite{MacWilliams:1977a, Steane:1999a} needs only a Hadamard gate to achieve universality, while the toric codes \cite{Kitaev:1997a} need both the $T$ gate and the Hadamard gate. As a variant of this approach, we could also use codes for which the remaining non-transversal gates have a low overhead magic state implementation. For example, the Hadamard gate magic state protocol likely requires much less overhead than the $T$ gate magic state protocol.

\item[(2)] The encoding magic states typically introduces significant noise. If this is our primary concern, we can try to find procedures for distilling magic states that allow for very noisy magic states as input. There are known theoretical bounds for this approach. Magic state distillation protocols use only Clifford circuits, and therefore states within the stabilizer polyhedron can never be distilled to non-stabilizer states. This is simply because the stabilizer states are closed under Clifford operations. There are, however, other conditions which preclude non-stabilizer states from being distillable, such as positivity of the discrete Wigner function \cite{Veitch2012}. Also the existence of bound states \cite{Campbell:2010a, Campbell:2009a, Campbell:2009b, Campbell:2010b} (states that cannot be distilled) have also been discovered. Nevertheless certain states that lie on the border of the stabilizer polyhedron have been shown to be distillable \cite{Reichardt:2005a}. Finding more states such as these should be our primary goal if we expect the initial magic states to be very noisy. The schemes discussed in this paragraph are illustrated in Fig.~\ref{fig:protocolunencoded}. We input some noisy states $\ket{\tilde{\rho}}$ into the distillation circuit (a). The approach outlined above seeks to find circuits (a) that allow very noisy inputs while still eventually distilling $\ket{M}$. If the system is noisy but qubits are abundant, this would be the appropriate paradigm to study. 

\item[(3)] If it is not difficult to prepare noisy magic states that meet the criteria for distill-ability, then our primary concern is to reduce overhead. Two ideas for reducing overhead are discussed below. Note that (3a) and (3b) are not mutually exclusive.

\begin{description}
\item[(3a)] For each magic state that we apply, we must first distill a high fidelity version of that magic state. This involves using many lower fidelity magic states as input to a distillation circuit with the goal of producing a higher fidelity magic state as output. Typically this process must be repeated for many rounds, feeding the output of one distillation circuit into the input of the next. The methods for reducing overhead in the distillation circuit involve either reducing the number of inputs needed at a given round or reducing the total number of rounds. Graphically, this approach focuses on improving the circuit in Fig.~\ref{fig:protocolunencoded}(a) by reducing the number of inputs and/or rounds of distillation. Protocols to reduce the number of rounds in magic state distillation were discussed in \cite{Meier:2012a}. 
	
\item[(3b)] Another way of reducing the overhead is to reuse magic states. This is accomplished by modifying the gate teleportation procedure such that the magic state is available for reuse after the gate has been applied. We will refer to these magic states as reusable magic states. This approach would allow magic states $\ket{M}$ that are input into the box in Fig.~\ref{fig:protocolunencoded}(b) to be reused without any additional distillation. If a code could be found such that its universal gate set was comprised of only transversal gates and reusable magic states, the savings in overhead would be immense. Once these magic states have been distilled there would be no need for any additional overhead ever! We will show that in most cases these reusable magic states are highly unlikely to exist. 		
\end{description}

\end{description}

\section{Reusable Magic States}
A quantum computing architecture that uses magic states consists of an encoded system $\mathcal{S}$ and a supply of encoded magic states in an auxiliary system $\mathcal{M}$. Here $\mathcal{M}$ refers to a fixed size, but otherwise arbitrary system. These systems should remain isolated until a magic state is needed in the computation. It is in this paradigm that we hope to implement gates with reusable magic states. 

Below we will represent our system simply as $\ket{\psi}$. We will represent the auxiliary system containing the magic state as $\ket{M}$. The argument that follows applies to both single qubit and encoded qubit systems. Also, these states can be mixed or pure. To make notation simpler, we will represent the systems as pure states on single qubit systems.

Formally we define a {\it reusable magic state} as a state $\ket{M}$ such that after application of a Clifford circuit on the joint system $\mathcal{S}\otimes \mathcal{M}$ some gate $U_{M}$ has been applied to the system $\mathcal{S}$ and the state $\ket{M}$ of the system $\mathcal{M}$ is unchanged. The state $\ket{M}$ can therefore be used again. We will use this definition to prove that reusable magic states do not exist for non-Clifford gates. When defining reusable magic states for Clifford gates the above definition must be restricted to prevent Clifford gates of the type are attempting to implement re-usably. 

A reusable magic state for the $S$ gate ($\sqrt{Z}$ gate) was shown in \cite{Aliferis:2007b, Jones:2010a}. The $S$ gate is a Clifford gate; however the circuit uses only $CNOT$ and $H$ to implement the $S$ gate.

\begin{figure}
\begin{equation*}
\Qcircuit @C=1.5em @R=1.5em {
 \ket{\pi/2} & & \targ        & \gate{H} & \targ      & \gate{H}  & \qw & \ket{\pi/2} \\
 \ket{\psi}  & & \ctrl{-1}   & \qw          & \ctrl{-1} & \qw  & \qw & S\ket{\psi}
 }
\end{equation*}
\caption{Gate teleportation circuit for the reusable S gate. The $\ket{\pi/2}$ magic state can be reused, thereby reducing the overhead for the $S$ gate to $\bigO(1)$. This circuit can be used in codes where $H$ is transversal, but $S$ is not. This particular circuit is credited to Austin Fowler and modified from \cite{Aliferis:2007b, Jones:2010a}.} 
\label{fig:teleportation4}
\end{figure}
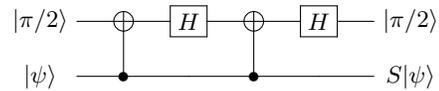

This gate can be modified to make a transversal $\sqrt{X}$ gate with the identity $\sqrt{X}=HSH$. Additionally, the combination of two reusable gates is itself a reusable gate. A reusable $\sqrt{Y}$ gate can be constructed by combining $\sqrt{X}$ and $S (\sqrt{Z})$ gates. It may seem that a reusable $H$ gate could be built through similar constructions; however all attempts by the author to date require that the $H$ gate be present in the circuit. This could only be the case when that gate is already available transversally, obviating the need for such a reusable magic state.

\section{Non-Clifford Reusable Magic States}
Most research on magic states focuses exclusively on non-Clifford magic states. In many technologies Clifford gates are considered to be easier to implement than general unitaries. They can be made universal with the addition of a single non-Clifford unitary. In other words, the gate set $<\{\mbox{Clifford}\},\; U>$ provides a dense set of unitaries in $SU(2^n)$. The canonical choice for $U$ is the $T$ gate ($\sqrt{S}$ gate). 

\begin{figure}
\begin{equation*}
\Qcircuit @C=1.5em @R=1.5em {
 \ket{\psi} & & \ctrl{1}  & \qw & \gate{S}  & \qw & T\ket{\psi} \\
 \ket{\pi/4} & & \targ   & \measureD{Z} & \control \cw  \cwx &
 }
\end{equation*}
\caption{Gate teleportation circuit for the T gate. In this circuit the magic state $\ket{\pi/4}$ is used to apply the T gate to some state $\ket{\psi}$. Where $\ket{\pi/4}=(\ket{0}+e^{i\pi/4}\ket{1})/\sqrt{2}$.} 
\label{fig:teleportation2}
\end{figure}
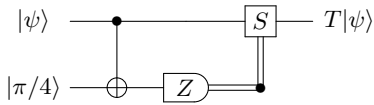

It is well-known that Clifford gates can be efficiently simulated on a classical computer in polynomial time {\bf P}. In fact, the computational power of a Clifford gate computer is thought to be weaker than a polynomial-time classical computer. 

The power of a polynomial-sized universal quantum gate set is, by definition, the class {\bf BQP} and hence can solve any problem within this class. Using the Solvay-Kitaev algorithm \cite{Kitaev:2002a, Nielsen:2000a}, we can compile any gate from a universal gate set in time linear in $\log^{c}(1/\epsilon)$. Where $c$ is some constant (typically between 2 and 3), and $\epsilon$ is the desired precision of the compiled gate. While some gate sets may be more efficient (in terms of overhead) than others, any universal quantum gate set can be used to efficiently solve problems in the class {\bf BQP}.  

In the derivation below we will present a `proof by contradiction'. We will assume the existence of a non-Clifford reusable magic state circuit, and then show that if such a circuit could be constructed, it would imply that {\bf BQP} $=$ {\bf P}. 

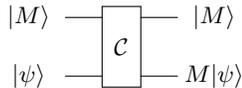
\begin{figure}
\begin{equation*}
\Qcircuit @C=1.5em @R=1.5em {
 \ket{M}    &  & \multigate{1}{\mathcal{C}} & \qw & \ket{M}\\
 \ket{\psi} &  & \ghost{\mathcal{C}} & \qw & M \ket{\psi}
 }
\end{equation*}
\caption{Reusable magic state circuit. $\mathcal{C}$ denotes some Clifford circuit and $\ket{M}$ is the reusable magic state. $\ket{M}$ may be comprised of many qubits and/or qudits providing the size is fixed. $M$ is any non-Clifford unitary. }
\label{fig:reusableMS}
\end{figure}

First, assume that the following circuit exists (see Fig.~\ref{fig:reusableMS}).

Where $\mathcal{C}$ denotes some Clifford circuit and $\ket{M}$ is the reusable magic state. $\ket{M}$ may be comprised of many qubits and/or qudits as long as the size is fixed. $M$ is any non-Clifford unitary.   

Now, since $<\{\mbox{Clifford}\},\; M>$ constitutes a universal gate set, we can write a general quantum circuit using only Clifford gates and $M$.
 
 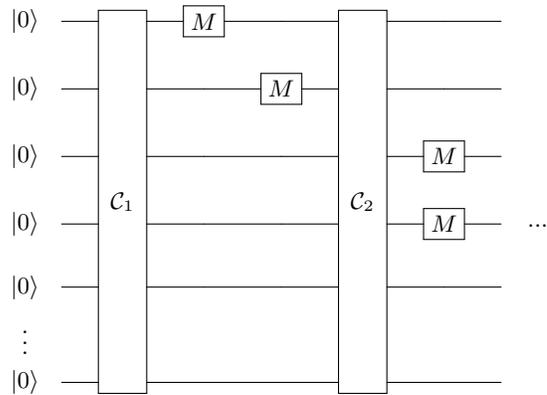
\begin{figure}
\begin{equation*}
\Qcircuit @C=1.5em @R=1.5em {
 \ket{0} &  & \multigate{6}{\mathcal{C}_1} & \gate{M} & \qw & \multigate{6}{\mathcal{C}_2} & \qw & \qw \\
 \ket{0} &  & \ghost{\mathcal{C}_1} & \qw & \gate{M}  & \ghost{\mathcal{C}_2} & \qw & \qw \\
 \ket{0} &  & \ghost{\mathcal{C}_1} & \qw & \qw & \ghost{\mathcal{C}_2} & \gate{M} & \qw \\
 \ket{0} &  & \ghost{\mathcal{C}_1} & \qw & \qw & \ghost{\mathcal{C}_2} & \gate {M} & \qw & ... \\
 \ket{0} &  & \ghost{\mathcal{C}_1} & \qw & \qw & \ghost{\mathcal{C}_2} & \qw & \qw & \\
 \vdots \\
 \ket{0} &  & \ghost{\mathcal{C}_1} & \qw & \qw & \ghost{\mathcal{C}_2} & \qw & \qw \\
 }
\end{equation*}
 \caption{ A generic quantum circuit. For clarity we have separated the Clifford gates into blocks, with the non-Clifford gate $M$ occurring between blocks. A general computation in {\bf BQP} would have polynomially many (in the number of inputs) rounds of Clifford and non-Clifford gates.}
 \label{fig:example}
 \end{figure}
 
For example, Fig.~\ref{fig:example} depicts a general quantum circuit with $\mathcal{C}_n$ denoting a round of arbitrary poly-sized Clifford gates, and $M$ a non-Clifford unitary. The circuit is simulate-able by a {\bf BQP} quantum computer as long as the circuit size (total number of circuit elements) is polynomial in the number of inputs. 

However, since we assumed that circuits of the form shown in Fig.~\ref{fig:reusableMS} exist, we can execute the same computation shown in Fig.~\ref{fig:example} by replacing the $M$ gates with their magic state implementation. Note that this will only increase the number of inputs by a constant amount (the size of $\ket{M}$). 

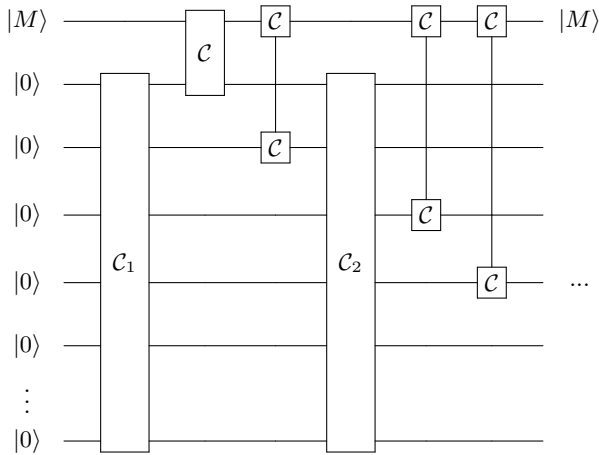
\begin{figure}
\begin{equation*}
\Qcircuit @C=1.5em @R=1.5em {
 \ket{M}&  & \qw                                         & \multigate{1}{\mathcal{C}} & \gate{\mathcal{C}} \qwx[1] & \qw & \gate{\mathcal{C}} \qwx[1] & \gate{\mathcal{C}} \qwx[1] & \qw &  \ket{M}\\
 \ket{0} &  & \multigate{6}{\mathcal{C}_1} &  \ghost{\mathcal{C}}            & \qw  \qwx[1]      & \multigate{6}{\mathcal{C}_2} & \qw \qwx[1] & \qw \qwx[1] & \qw & \\
 \ket{0} &  & \ghost{\mathcal{C}_1}             & \qw                                         &\gate{\mathcal{C}} & \ghost{\mathcal{C}_2}   & \qw \qwx[1] & \qw \qwx[1] & \qw & \\
 \ket{0} &  & \ghost{\mathcal{C}_1} & \qw & \qw & \ghost{\mathcal{C}_2} & \gate{\mathcal{C}} & \qw \qwx[1] & \qw & \\
 \ket{0} &  & \ghost{\mathcal{C}_1} & \qw & \qw & \ghost{\mathcal{C}_2} & \qw & \gate{\mathcal{C}} & \qw & ... \\
 \ket{0} &  & \ghost{\mathcal{C}_1} & \qw & \qw & \ghost{\mathcal{C}_2} & \qw & \qw & \qw & \\
 \vdots \\
 \ket{0} &  & \ghost{\mathcal{C}_1} & \qw & \qw & \ghost{\mathcal{C}_2} & \qw & \qw & \qw & \\
 }
\end{equation*}
\caption{The circuit in Fig.~\ref{fig:example} with all non-Clifford gates $M$ replaced with the circuit in Fig.~\ref{fig:reusableMS}. Now, all non-Clifford circuitry has been moved to the beginning of the computation, and only constant overhead (the size of $\ket{M}$) has been introduced.}
\label{fig:example2}
\end{figure}

We can continue this process and replace all gates $M$ by their magic state implementations. Now the entire body of the computation consists of only Clifford gates. We have only to prepare the state $\ket{M}$ which is unentangled with the rest of the system. This state could still be highly non-trivial; however we can always represent this state as a sum of stabilizer states. For example, the single qubit pure state $\ket{\psi} = \alpha\ket{0} + \beta\ket{1}$ can be written as $|\alpha|^{2}\ket{0}\bra{0} + \alpha\beta^{*}\ket{0}\bra{1} + \alpha^{*} \beta\ket{1}\bra{0} + |\beta|^{2}\ket{1}\bra{1} = a_{+}I + a_{-}Z + b_{+}X + (-i)b_{-}Y$. Where $a_{\pm} = \frac{|\alpha|^{2} \pm |\beta|^{2}}{2}$ and $b_{\pm} = \frac{\alpha\beta^{*} \pm \alpha^{*}\beta}{2}$. We have fixed the size of $\ket{M}$ to be independent of circuit size; therefore it can always assumed that the circuit size is large enough, such that the dimension of $\ket{M}$ is logarithmic in number of inputs to the circuit. We can then write $\ket{M}$ as a sum of stabilizers which will generally have a number of terms that grows as $\bigO(2^{d})$. Where $d$ is the dimension of $\ket{M}$. Again, this number is fixed and is independent of circuit size. This amounts to a constant overhead in our notation. Finally, since the entire body of the circuit consists of Clifford gates which map stabilizer states to stabilizer states, the number of terms in the initial sum of stabilizer states is fixed throughout the computation. We can simulate each of the terms in the initial sum of stabilizer states in time that grows polynomially with the number of input states. We can thus simulate the entire circuit in time $\bigO(2^{d}\times POLY(n)) = \bigO(POLY(n))$. Where $d=c$ (some constant) and $n$ is the number of input states. 

In conclusion, we have shown that if a circuit such as that shown in Fig.~\ref{fig:reusableMS} exists for non-Clifford unitary $M$, then {\bf BQP} $=$ {\bf P}. In fact, since Clifford state computation is in the class {\bf ParityL} \cite{Aaronson:2004a} (which is thought to be weaker than {\bf P}), this would be of even greater consequence. In the highly unlikely event that such a circuit exists, it would not be useful since the entire endeavor of quantum computation would be obviated as a consequence. 

Some open questions still linger such as: {\it Does a reusable magic state exist for the $H$ gate?} This circumvents the proof in this paper, since $H$ {\it is} a Clifford gate. Codes such as the 15-qubit Reed-Muller code can be made universal with the addition of such a gate; therefore finding such a state would drastically reduce the overhead for this and similar codes. As mentioned above, our definition of reusable magic states must be modified when the unitary we are trying to implement is a Clifford gate. 

Qudit magic state distillation has recently been introduced in \cite{Campbell:2012a, Anwar:2012a, Veitch2012}. Our result applies to qudit codes as well. Namely, non-Clifford qudit gates cannot be implemented using reusable magic states, unless qudit quantum computation is efficiently simulate-able on a classical computer. The proof is briefly sketched here: The Clifford group for any prime number $p$ in $SU(p^n)$ is a maximal finite subgroup. The addition of any non-Clifford unitary generates an infinite group which is dense in $SU(p^n)$. As in the qubit case, we need only a single non-Clifford gate to complete a universal gate set. These properties of the Clifford group are not well known and were only recently mentioned in the physics literature (see appendix in \cite{Campbell:2012a} and references therein). Using this, the proof for the qudit case follows in exactly the same manner as the qubit case. 

It is, however, possible that qudit analogues of the Reed-Muller or other similar codes can complete a universal gate set with the addition of some qudit Clifford gate. Therefore, it may be fruitful to search for these codes and for reusable qudit magic states.

\begin{acknowledgments}

The author would like to acknowledge many helpful conversations with Chris Cesare, Andrew Landahl, Rolando Somma, Adam Meier, Bryan Eastin, Jim Harrington, and Olivier Landon-Cardinal. Additionally, I would like to thank Earl Campbell for pointing out subtleties in the qudit Clifford group and providing ideas for extending this proof to the qudit case. 

JTA was supported
in part by the National Science Foundation through Grant 0829944. JTA was supported in part by the
Laboratory Directed Research and Development program at Sandia National
Laboratories.  Sandia National Laboratories is a multi-program laboratory
managed and operated by Sandia Corporation, a wholly owned subsidiary of
Lockheed Martin Corporation, for the U.S.  Department of Energy's  National
Nuclear Security Administration under contract DE-AC04-94AL85000.

\end{acknowledgments}

%


\begin{thebibliography}{20}
\expandafter\ifx\csname natexlab\endcsname\relax\def\natexlab#1{#1}\fi
\expandafter\ifx\csname bibnamefont\endcsname\relax
  \def\bibnamefont#1{#1}\fi
\expandafter\ifx\csname bibfnamefont\endcsname\relax
  \def\bibfnamefont#1{#1}\fi
\expandafter\ifx\csname citenamefont\endcsname\relax
  \def\citenamefont#1{#1}\fi
\expandafter\ifx\csname url\endcsname\relax
  \def\url#1{\texttt{#1}}\fi
\expandafter\ifx\csname urlprefix\endcsname\relax\def\urlprefix{URL }\fi
\providecommand{\bibinfo}[2]{#2}
\providecommand{\arxiv}[2][]{\href{http://arxiv.org/pdf/#2}{\texttt{arXiv:#2}}}
\providecommand{\doi}[2][]{\href{http://dx.doi.org/#2}{\texttt{doi:#2}}}

\bibitem[{\citenamefont{Bravyi and Kitaev}(2005)}]{Bravyi:2005a}
\bibinfo{author}{\bibfnamefont{S.}~\bibnamefont{Bravyi}} \bibnamefont{and}
  \bibinfo{author}{\bibfnamefont{A.}~\bibnamefont{Kitaev}},
  \emph{\bibinfo{title}{{Universal quantum computation with ideal \{C\}lifford
  gates and noisy ancillas}}}, \bibinfo{journal}{Phys. Rev. A}
  \textbf{\bibinfo{volume}{71}}, \bibinfo{pages}{22316} (\bibinfo{year}{2005}),
  \doi{10.1103/PhysRevA.71.022316}.

\bibitem[{\citenamefont{Eastin and Knill}(2008)}]{Eastin:2008a}
\bibinfo{author}{\bibfnamefont{B.}~\bibnamefont{Eastin}} \bibnamefont{and}
  \bibinfo{author}{\bibfnamefont{E.}~\bibnamefont{Knill}},
  \emph{\bibinfo{title}{{Restrictions on transversal encoded quantum gate
  sets}}} (\bibinfo{year}{2008}), \doi{10.1103/PhysRevLett.102.110502},
  \arxiv{0811.4262}.

\bibitem[{\citenamefont{Kitaev}(2003)}]{Kitaev:1997a}
\bibinfo{author}{\bibfnamefont{A.~Y.} \bibnamefont{Kitaev}},
  \emph{\bibinfo{title}{{Fault-tolerant quantum computation by anyons}}},
  \bibinfo{journal}{Ann. Phys.} \textbf{\bibinfo{volume}{303}},
  \bibinfo{pages}{2} (\bibinfo{year}{2003}),
  \doi{10.1016/S0003-4916(02)00018-0}.

\bibitem[{\citenamefont{Koenig et~al.}(2010)\citenamefont{Koenig, Kuperberg,
  and Reichardt}}]{Koenig:2010a}
\bibinfo{author}{\bibfnamefont{R.}~\bibnamefont{Koenig}},
  \bibinfo{author}{\bibfnamefont{G.}~\bibnamefont{Kuperberg}},
  \bibnamefont{and} \bibinfo{author}{\bibfnamefont{B.~W.}
  \bibnamefont{Reichardt}}, \emph{\bibinfo{title}{{Quantum computation with
  Turaev-Viro codes}}}, p.~\bibinfo{pages}{53} (\bibinfo{year}{2010}),
  \arxiv{1002.2816}, \urlprefix\url{http://arxiv.org/abs/1002.2816}.

\bibitem[{\citenamefont{MacWilliams and Sloane}(1977)}]{MacWilliams:1977a}
\bibinfo{author}{\bibfnamefont{F.~J.} \bibnamefont{MacWilliams}}
  \bibnamefont{and} \bibinfo{author}{\bibfnamefont{N.~J.~A.}
  \bibnamefont{Sloane}}, \emph{\bibinfo{title}{{The Theory of Error-Correcting
  Codes}}}, vol.~\bibinfo{volume}{16} of \emph{\bibinfo{series}{North-Holland
  mathematical library}} (\bibinfo{publisher}{North-Holland},
  \bibinfo{address}{New York}, \bibinfo{year}{1977}), ISBN
  \bibinfo{isbn}{0-444-85009-0}.

\bibitem[{\citenamefont{Steane}(1999)}]{Steane:1999a}
\bibinfo{author}{\bibfnamefont{A.}~\bibnamefont{Steane}},
  \emph{\bibinfo{title}{{Quantum Reed-Muller codes}}}, \bibinfo{journal}{IEEE
  Transactions on Information Theory} \textbf{\bibinfo{volume}{45}},
  \bibinfo{pages}{1701} (\bibinfo{year}{1999}), ISSN \bibinfo{issn}{00189448},
  \doi{10.1109/18.771249},
  \urlprefix\url{http://ieeexplore.ieee.org/lpdocs/epic03/wrapper.htm?arnumber%
=771249}.

\bibitem[{\citenamefont{Veitch et~al.}(2012)\citenamefont{Veitch, Ferrie, and
  Emerson}}]{Veitch2012}
\bibinfo{author}{\bibfnamefont{V.}~\bibnamefont{Veitch}},
  \bibinfo{author}{\bibfnamefont{C.}~\bibnamefont{Ferrie}}, \bibnamefont{and}
  \bibinfo{author}{\bibfnamefont{J.}~\bibnamefont{Emerson}},
  \emph{\bibinfo{title}{{Negative Quasi-Probability Representation is a
  Necessary Resource for Magic State Distillation}}},
  \bibinfo{journal}{Arxiv:1201.1256}  (\bibinfo{year}{2012}).

\bibitem[{\citenamefont{Campbell and Browne}(2010)}]{Campbell:2010a}
\bibinfo{author}{\bibfnamefont{E.~T.} \bibnamefont{Campbell}} \bibnamefont{and}
  \bibinfo{author}{\bibfnamefont{D.~E.} \bibnamefont{Browne}},
  \emph{\bibinfo{title}{{Bound States for Magic State Distillation in
  Fault-Tolerant Quantum Computation}}}, \bibinfo{journal}{Phys. Rev. Lett.}
  \textbf{\bibinfo{volume}{104}} (\bibinfo{year}{2010}).

\bibitem[{\citenamefont{Campbell and
  Browne}(2009{\natexlab{a}})}]{Campbell:2009a}
\bibinfo{author}{\bibfnamefont{E.~T.} \bibnamefont{Campbell}} \bibnamefont{and}
  \bibinfo{author}{\bibfnamefont{D.~E.} \bibnamefont{Browne}},
  \emph{\bibinfo{title}{{On the Structure of Protocols for Magic State
  Distillation}}}, \bibinfo{journal}{Lecture Notes in Computer Science 5906
  Theory of Quantum Computation, Communication and Cryptography}
  p.~\bibinfo{pages}{20} (\bibinfo{year}{2009}{\natexlab{a}}).

\bibitem[{\citenamefont{Campbell and
  Browne}(2009{\natexlab{b}})}]{Campbell:2009b}
\bibinfo{author}{\bibfnamefont{E.~T.} \bibnamefont{Campbell}} \bibnamefont{and}
  \bibinfo{author}{\bibfnamefont{D.~E.} \bibnamefont{Browne}},
  \emph{\bibinfo{title}{{Bound States for Magic State Distillation}}}
  (\bibinfo{year}{2009}{\natexlab{b}}).

\bibitem[{\citenamefont{Campbell}(2010)}]{Campbell:2010b}
\bibinfo{author}{\bibfnamefont{E.~T.} \bibnamefont{Campbell}},
  \emph{\bibinfo{title}{{Catalysis and activation of magic states in fault
  tolerant architectures}}} (\bibinfo{year}{2010}),
  \doi{10.1103/PhysRevA.83.032317}, \arxiv{1010.0104},
  \urlprefix\url{http://arxiv.org/abs/1010.0104}.

\bibitem[{\citenamefont{Reichardt}(2005)}]{Reichardt:2005a}
\bibinfo{author}{\bibfnamefont{B.~W.} \bibnamefont{Reichardt}},
  \emph{\bibinfo{title}{{Quantum Universality from Magic States Distillation
  Applied to \{CSS\} Codes}}}, \bibinfo{journal}{Quant. Inf. Proc.}
  \textbf{\bibinfo{volume}{4}}, \bibinfo{pages}{251} (\bibinfo{year}{2005}),
  \doi{10.1007/s11128-005-7654-8}.

\bibitem[{\citenamefont{Meier et~al.}(2012)\citenamefont{Meier, Eastin, and
  Knill}}]{Meier:2012a}
\bibinfo{author}{\bibfnamefont{A.~M.} \bibnamefont{Meier}},
  \bibinfo{author}{\bibfnamefont{B.}~\bibnamefont{Eastin}}, \bibnamefont{and}
  \bibinfo{author}{\bibfnamefont{E.}~\bibnamefont{Knill}},
  \emph{\bibinfo{title}{{Magic-state distillation with the four-qubit code}}},
  p.~\bibinfo{pages}{10} (\bibinfo{year}{2012}), \arxiv{1204.4221},
  \urlprefix\url{http://arxiv.org/abs/1204.4221}.

\bibitem[{\citenamefont{Aliferis}(2007)}]{Aliferis:2007b}
\bibinfo{author}{\bibfnamefont{P.}~\bibnamefont{Aliferis}},
  \emph{\bibinfo{title}{{Level reduction and the quantum threshold theorem}}},
  Ph.D. thesis, \bibinfo{school}{Caltech} (\bibinfo{year}{2007}).

\bibitem[{\citenamefont{Jones et~al.}(2010)\citenamefont{Jones, {Van Meter},
  Fowler, McMahon, Kim, Ladd, and Yamamoto}}]{Jones:2010a}
\bibinfo{author}{\bibfnamefont{N.~C.} \bibnamefont{Jones}},
  \bibinfo{author}{\bibfnamefont{R.}~\bibnamefont{{Van Meter}}},
  \bibinfo{author}{\bibfnamefont{A.~G.} \bibnamefont{Fowler}},
  \bibinfo{author}{\bibfnamefont{P.~L.} \bibnamefont{McMahon}},
  \bibinfo{author}{\bibfnamefont{J.}~\bibnamefont{Kim}},
  \bibinfo{author}{\bibfnamefont{T.~D.} \bibnamefont{Ladd}}, \bibnamefont{and}
  \bibinfo{author}{\bibfnamefont{Y.}~\bibnamefont{Yamamoto}},
  \emph{\bibinfo{title}{{Layered architecture for quantum computing}}},
  p.~\bibinfo{pages}{23} (\bibinfo{year}{2010}), \arxiv{1010.5022},
  \urlprefix\url{http://arxiv.org/abs/1010.5022}.

\bibitem[{\citenamefont{Kitaev et~al.}(2002)\citenamefont{Kitaev, Shen, and
  Vyalyi}}]{Kitaev:2002a}
\bibinfo{author}{\bibfnamefont{A.~b.~Y.} \bibnamefont{Kitaev}},
  \bibinfo{author}{\bibfnamefont{A.}~\bibnamefont{Shen}}, \bibnamefont{and}
  \bibinfo{author}{\bibfnamefont{M.~N.} \bibnamefont{Vyalyi}},
  \emph{\bibinfo{title}{{Classical and Quantum Computation}}},
  vol.~\bibinfo{volume}{47} of \emph{\bibinfo{series}{Graduate Studies in
  Mathematics}} (\bibinfo{publisher}{American Mathematical Society},
  \bibinfo{address}{Providence, RI}, \bibinfo{year}{2002}), ISBN
  \bibinfo{isbn}{0-821-82161-X}.

\bibitem[{\citenamefont{Nielsen and Chuang}(2000)}]{Nielsen:2000a}
\bibinfo{author}{\bibfnamefont{M.~A.} \bibnamefont{Nielsen}} \bibnamefont{and}
  \bibinfo{author}{\bibfnamefont{I.~L.} \bibnamefont{Chuang}},
  \emph{\bibinfo{title}{{Quantum Computation and Quantum Information}}}
  (\bibinfo{publisher}{Cambridge University Press},
  \bibinfo{address}{Cambridge}, \bibinfo{year}{2000}), ISBN
  \bibinfo{isbn}{0-521-63235-8 (Hardback), 0-521-63503-9 (Paperback)}.

\bibitem[{\citenamefont{Aaronson and Gottesman}(2004)}]{Aaronson:2004a}
\bibinfo{author}{\bibfnamefont{S.}~\bibnamefont{Aaronson}} \bibnamefont{and}
  \bibinfo{author}{\bibfnamefont{D.}~\bibnamefont{Gottesman}},
  \emph{\bibinfo{title}{{Improved simulation of stabilizer circuits}}},
  \bibinfo{journal}{Phys. Rev. A} \textbf{\bibinfo{volume}{70}},
  \bibinfo{pages}{52328} (\bibinfo{year}{2004}),
  \doi{10.1103/PhysRevA.70.052328}.

\bibitem[{\citenamefont{Campbell et~al.}(2012)\citenamefont{Campbell, Anwar,
  and Browne}}]{Campbell:2012a}
\bibinfo{author}{\bibfnamefont{E.~T.} \bibnamefont{Campbell}},
  \bibinfo{author}{\bibfnamefont{H.}~\bibnamefont{Anwar}}, \bibnamefont{and}
  \bibinfo{author}{\bibfnamefont{D.~E.} \bibnamefont{Browne}},
  \emph{\bibinfo{title}{{Magic state distillation in all prime dimensions using
  quantum Reed-Muller codes}}} (\bibinfo{year}{2012}), \arxiv{1205.3104},
  \urlprefix\url{http://arxiv.org/abs/1205.3104}.

\bibitem[{\citenamefont{Anwar et~al.}(2012)\citenamefont{Anwar, Campbell, and
  Browne}}]{Anwar:2012a}
\bibinfo{author}{\bibfnamefont{H.}~\bibnamefont{Anwar}},
  \bibinfo{author}{\bibfnamefont{E.~T.} \bibnamefont{Campbell}},
  \bibnamefont{and} \bibinfo{author}{\bibfnamefont{D.~E.}
  \bibnamefont{Browne}}, \emph{\bibinfo{title}{{Qutrit Magic State
  Distillation}}}, p.~\bibinfo{pages}{13} (\bibinfo{year}{2012}),
  \arxiv{1202.2326}, \urlprefix\url{http://arxiv.org/abs/1202.2326}.

\end{thebibliography}
\end{document}